\documentclass[rmp,12pt]{revtex4}
\usepackage{epsfig}
\usepackage{bm}
\usepackage{amsmath}
\usepackage{enumitem}
\usepackage{color}
\usepackage[colorlinks=true,pdfstartview=FitV,linkcolor=blue,citecolor=blue,urlcolor=blue]{hyperref}
\begin{document}
\vspace{1cm}
\def\thesubsection{\arabic{subsection}}

\def\HPs{$(H,\Psi)$ }
\def\HP{$(H,\Psi)$}
\title{\Huge Essays}
\author{James B.~Hartle}

\email{hartle@physics.ucsb.edu}

\affiliation{Department of Physics, University of California,Santa Barbara, CA 93106-9530}
\affiliation{Santa Fe Institute, Santa Fe, NM 87501}

\date{\today}

\begin{abstract} 
 \normalsize A collection of short expository essays on various topics in quantum mechanics, quantum cosmology, and physics in general. 
\end{abstract}

\maketitle

\section*{\large Introduction} 
From time to time  the author has had  an  occasion or an impulse to write a short piece of an essentially expository nature. These essays mostly address issues that arise frequently in physics --- for example those connected with understanding quantum mechanics.  A number of these essays are collected here. 

The essays are  are written at a level that should be accessible to  many physicists, but they are not popular articles aimed at a general public. They do not replace the author's longer efforts at exposition for example those in his Jerusalem Lectures \href{http:arxiv.org/abs/1805.12246}{(91)} and Les Houches lectures \href{http:arxiv.org/abs/gr-qc/9304006}{(107)}. 

The essays were written over a period of thirty years and vary significantly in length, style, subject, quality, level, overlap, point of view, and perhaps even consistency between them.

\section*{\large Organization}

Each essay listed below has a title followed a brief description, sometimes just the abstract but other times a shorter description.
Following the title there is a clickable link to the full paper on arXiv.

The essays are divided into three rough catagories:
\begin{itemize}
\item{\it  General}: Having something to do with physical theories generally.
\item{\it Quantum Mechanics:}  Dealing first with issues in standard Schroedinger-Heisenberg quantum mechanics of a closed system in a fixed background spacetime. 
\item{\it Cosmology:} Dealing with aspects of the quantum  universe,  in particular with the generalizations of standard quantum mechanics necessary for quantum spacetime geometry (quantum gravity) and  quantum cosmology. 

\end{itemize}
\noindent Within each category the papers are ordered  by date of appearance. {\bf When a paper seems to fall into two catagories its description appears in both. }

\section*{\large General}
 
\subsection{Computability and Physical Theory   \href{http:arxiv.org/abs/1806.09237}{(85)}}
\noindent
(with Robert Geroch) What would physics be like if its theories predicted measurable numbers that were non-computable in the mathematical sense? That is, numbers for which no one computer program can exist to compute them to any accuracy that may be specified. Not that much would be changed, but existence of algorithms for approximate computation would be an important question for such theories. 

\subsection{Excess Baggage  \href{http:arxiv.org/abs/gr-qc/0508001}{(85)}}
In many advances in physics some previously accepted general idea was found to be unnecessary and dispensable. The idea was not truly a general feature of the world, but only perceived to be general because of our special place in the universe and the limited range of our experience. It was excess baggage which had to be jettisoned to reach a more a more general perspective. This article discusses excess baggage from the perspective of quantum cosmology, and asks what is excess baggage in our present theoretical framework?

\subsection{Sources of Predictability,  \href{http:arxiv.org/abs/gr-qc/9701027}{(112)}} 
Sources of predictability from the basic laws of physics are described in the most general theoretical context --- the quantum theory of the universe.

\subsection{Quantum Pasts and the Utility of History, \href{http:arxiv.org/abs/gr-qc/9712001}{(117)}}
Why are we interested in the past? Its over and done with. This article describes the process of retrodicting the past in quantum cosmology using quantum probabilities conditioned on present data. There is not just one past but many different possible represented by different decohering sets of alternative coarse-grained histories of the past.  Retrodicting a past is useful because it can help with predicting the future. 

\subsection{Theories of Everything and Hawking's Wave Function of the Universe, \href{http:arxiv.org/abs/gr-qc/0209047}{(125)}}
If a cat, a cannonball, and an economic textbook are dropped from a building they all fall to the ground with the same acceleration. That is an example of a universal law of physics. But that fact tells us little about cats, canonballs, or economics. This essay describes how a theory of these universal laws --- what in physics is called `a theory of everything' --- must include a quantum state of the universe. A `theory of everything' in fact does not predict everything we observe but only certain features of what we observe that are related to the universal regularities exhibited by all physical systems without exception, without qualification, and without approximation. 

\subsection{Scientific Knowledge from the Perspective of Quantum Cosmology, \href{http:arxiv.org/abs/gr-qc/9601046}{(108)}}
Scientific knowledge is limited for at least three reasons: (1) Physical theories predict only the regularities of our experience and not every detail of it. (2)  Predictions may require intractable computation (3) The process of induction and test is limited. Quantum cosmology displays all three kinds of limits. This essay briefly describes them and the place of the other sciences in this most comprehensive of physical frameworks. 

\subsection{The Physics of Now, \href{http:arxiv.org/abs/gr-qc/0403001}{(31)}}
The world is four-dimensional according to fundamental physics, governed by basic laws that operate in a spacetime that has no unique division into space and time. Yet our subjective experience is divided into present, past, and future. This paper discusses the origin of this division in terms of simple models of information gathering and utilizing systems (IGUSes). Past, present, and future are not properties of four-dimensional spacetime but notions describing how individual IGUSes process information. Their origin is to be found in how these IGUSes evolved or were constructed. The past, present, and future of an IGUS is consistent with the four-dimensional laws of physics and can be described in four-dimensional terms. The present  is not a moment of time in the sense of a spacelike surface in spacetime. Rather there is a localized notion of present at each point along an IGUS' world line. The common present of a group of  localized IGUSes is an approximate notion appropriate when they are sufficiently close to each other and have relative velocities much less than that of light. Modes of temporal organization that are different from present, past and future can be imagined that are consistent with the physical laws. We speculate why the present, past, and future organization might be favored by evolution and therefore a cognitive universal to be found here on Earth and other places in the Universe.



\section*{\large Quantum Mechanics} 

\subsection{The Quantum Mechanics of Closed Systems \href{http:arxiv.org/abs/gr-qc/9210006}{(98)}}
A pedagogical introduction is given to the quantum mechanics of closed systems, most generally the universe as a whole. Quantum mechanics aims at predicting the probabilities of alternative coarse-grained time histories of a closed system. But, not every set of alternative coarse-grained histories that can be described may be consistently assigned probabilities because of quantum mechanical interference between individual histories of the set. 

In ÒCopenhagenÓ quantum mechanics, probabilities can be assigned to histories of a subsystem that have been ÒmeasuredÓ. In the quantum mechanics of closed systems, containing both observer and observed, probabilities are assigned to those sets of alternative histories for which there is negligible interference between individual histories as a consequence of the systemÕs initial condition and dynamics. Such sets of histories are said to decohere. We define decoherence for closed systems in the simplified case when quantum gravity can be neglected and the initial state is pure. 

Copenhagen quantum mechanics is an approximation to the more general quantum framework of closed systems. It is an approximation that is appropriate when there is an approximately isolated subsystem that is a participant in a measurement situation in which (among other things) the decoherence of alternative registrations of the apparatus can be idealized as exact.

\subsection{The Spacetime Approach to Quantum Mechanics \href{http:arxiv.org/abs/gr-qc/9210004}{(99)}}
Feynman's sum-over-histories formulation of quantum mechanics is reviewed as an independent statement of quantum theory in spacetime form. It is different from the usual Schroedinger-Heisenberg formulation that utilizes states on spacelike surfaces because it assigns probabilities to different sets of alternatives. 
The general notion of a set of spacetime alternatives is a partition (coarse-graining) of the histories into an exhaustive set of exclusive classes. With this generalization the sum-over-histories formulation can be said to be in fully spacetime form with dynamics represented by path integrals over spacetime histories and alternatives defined as spacetime partitions of these histories. When restricted to alternatives at definite moments of times this generalization is equivalent to Schroedinger-Heisenberg quantum mechanics. However, the quantum mechanics of more general spacetime alternatives does not have an equivalent Schroeodinger-Heisenberg formulation. We suggest that, in the quantum theory of gravity, the general notion of ÒobservableÓ is supplied by diffeomorphism invariant partitions of spacetime metrics and matter field configurations. By generalizing the usual alternatives so as to put quantum theory in fully spacetime form we may be led to a covariant generalized quantum mechanics of spacetime free from the problem of time.

\subsection{Quantum Physics and Human Language   \href{http:arxiv.org/abs/quant-ph/0610131}{(136)}}
Human languages tacitly assume specific properties of the limited world they evolved to describe. These properties are true features of that limited context, but may not be general or precise properties in fundamental physics. Human languages must therefore be qualified, discarded, or otherwise reformed to give a clear account from fundamental physics of even the phenomena that the languages evolved to describe. The surest route to clarity is to express the constructions of human languages in the language of fundamental physical theory, not the other way around. These ideas are illustrated by an analysis of the verb `to happen' and the word `reality' in special relativity and the modern quantum mechanics of closed systems. This paper contains the author's views on what is `real'. 

\subsection{Quantum Pasts and the Utility of History \href{http:arxiv.org/abs/gr-qc/9712001}{(117)}}
Why are we interested in the past? Its over and done with. This article describes the process of retrodicting the past in quantum cosmology with quantum probabilities conditioned on present data. There is not just one past but many different possible ones corresponding to different decohering sets of alternative coarse-grained past histoies. Retrodicting a past is useful because helps with predicting the future. 

\subsection{The Quasiclassical Realms of this Quantum Universe \href{http:arxiv.org/abs/0806.3776}{(141)}}
The most striking observable feature of our indeterministic quantum universe is the wide range of time, place, and scale on which the deterministic laws of classical physics hold to an excellent approximation. This essay describes how this domain of classical predictability of every day experience emerges from a quantum theory of the universeÕs state and dynamics.

\subsection{Living in a Superposition \href{http:arxiv.org/abs/1511.01550}{(159)}}
This essay considers a model quantum universe consisting of a very large box containing a screen with two slits and an observer (us) that can pass though the slits. We apply the modern quantum mechanics of closed systems to calculate the probabilities for alternative histories of how we move through this universe and what we see. After passing through the screen with the slits, the quantum state of the universe is a superposition of classically distinguishable histories. We are then living in a superposition. Some frequently asked questions about such situations are answered using this model. The modelÕs relationship to more realistic quantum cosmologies is briefly discussed.

\subsection{The Quantum Mechanical Arrows of Time \href{http:arxiv.org/abs/1301.2844}{(150)}}
The familiar textbook quantum mechanics of laboratory measurements incorporates a quantum mechanical arrow of time --- the direction in time in which state vector reduction operates. This arrow is usually assumed to coincide with the direction of the thermodynamic arrow of the quasiclassical realm of everyday experience. But in the more general context of cosmology we seek an explanation of all observed arrows, and the relations between them, in terms of the conditions that specify our particular universe. This essay investigates quantum mechanical and thermodynamic arrows in a time-neutral formulation of quantum mechanics for a number of model cosmologies in fixed background spacetimes. We find that a general universe may not have well defined arrows of either kind. When arrows are emergent they need not point in the same direction over the whole of spacetime. Rather they may be local, pointing in different directions in different spacetime regions. Local arrows can therefore be consistent with global time symmetry.

\subsection{Why our Universe is Comprehensible  \href{http:arxiv.org/abs/1612.01952}{(162)}}
Einstein wrote memorably that `The eternally incomprehensible thing about the world is its comprehensibility.Õ This essay argues that the universe must be comprehensible for information gathering and utilizing systems such as human observers to evolve and function.

\subsection{The Impact  of Cosmology on Quantum Mechanics \href{http:arxiv.org/abs/1901.03933}{(173)}}

When quantum mechanics was developed in the '20s of the last century another revolution in physics was just starting. It began with the discovery that the universe is expanding. For a long time quantum mechanics and cosmology developed  independently of one another. Yet the very discovery of the expansion would eventually draw the two subjects together because it implied the big bang where quantum mechanics was important for cosmology and for understanding and predicting our observations of the universe today. 
Textbook (Copenhagen)  formulations of quantum mechanics are inadequate for cosmology for at least four reasons: 1) They predict  the outcomes of measurements made by observers. But in the very early  universe no measurements were being made and no observers  were around to make them. 2) Observers were outside of the system being measured.  But we are interested in a theory of the whole universe where everything, including observers, are inside. 3) Copenhagen quantum mechanics could not retrodict the past. But retrodicting the past  to understand how the universe began is the main  task of cosmology. 4) Copenhagen quantum mechanics  required a  fixed classical spacetime geometry not least to give meaning to the  time in the Schr\"odinger equation. But in the very early universe spacetime is fluctuating quantum mechanically (quantum gravity) and without definite value. 
A formulation of quantum mechanics general enough for cosmology was started by Everett and developed by many. 
That effort has given us a more general  framework that is adequate for cosmology --- decoherent (or consistent) histories quantum theory in the context of semiclassical quantum gravity.  Copenhagen quantum theory is an approximation to this more general quantum framework that is appropriate for measurement situations.  We discuss whether further generalization may still be required.

\section*{\large Cosmology}

\subsection{Quantum Cosmology: Problems for the 21st Century \href{http:arxiv.org/abs/gr-qc/9701022}{(113)}}
Two fundamental laws are needed for prediction in the universe: (1) a basic dynamical law and (2) a law for the quantum state of the universe.. Quantum cosmology is the area of basic research concerned with the search for a theory of the initial cosmological state. The issues involved in this search are presented in the form of eight problems.

\subsection{Anthropic Reasoning and Quantum Cosmology \href{http:arxiv.org/abs/gr-qc/0406104}{(129)}} 
In quantum cosmology anthropic reasoning uses probabilities that are conditioned on the existence of us as physical observing systems within the universe. This essay discusses how anthropic reasoning depends on the quantum state of the universe. Every prediction for our observations of the universe involves anthropic considerations.

\subsection{Scientific Knowledge from the Perspective of Quantum Cosmology \href{http:arxiv.org/abs/gr-qc/9601046}{(108)}} 
Scientific knowledge is limited for at least three reasons: Physical theories predict only the regularities of our experience and not every detail of it, predictions may require intractable computation, and the process of induction and test is limited. Quantum cosmology displays all three kinds of limits. This essay briefly describes them and the place of the other sciences in this most comprehensive of physical frameworks. 

\subsection {Are We Typical?  \href{http:arxiv.org/abs/0704.2630}{(137)}}
(with Mark Srednicki.) We have no observational evidence that as human observers we are typical of any class of objects in the universe, and there is no reason to believe that the laws of physics have to be such as to make our observations typical of others that might be made in the universe. Indeed an assumption that we are atypical is a testable hypothesis.

\subsection{Science in a Very Large Universe \href{http:arxiv.org/abs/0906.0042}{(144)},
\href{http:arxiv.org/abs/1004.3816}{(144A)}}
(with Mark Srednicki) Inflation can make the universe large enough that there is significant probability that we are replicated as physical systems at other locations in the universe. Predictions of our future observations then require an assumed probability distribution for our location among the possible ones (the xerographic distribution) in addition to the probabilities arising from the quantum state. It is the combination of fundamental theory plus the xerographic distribution that can be predictive and testable by further observations.

\subsection{The Quasiclassical Realms of this Quantum Universe   \href{http:arxiv.org/abs/1801.08631}{(141)}}
In this universe, governed fundamentally by quantum mechanical laws, characterized by indeterminism and distributed probabilities, classical deterministic laws are applicable over a wide range of time, place, and scale. We review the origin of these deterministic laws in the context of the quantum mechanics of closed systems, most generally, the universe as a whole. In this formulation of quantum mechanics, probabilities are predicted for the individual members of sets of alternative coarse-grained histories of the universe that decohere, i.e., for which there is negligible interference between pairs of histories in the set as measured by a decoherence functional. 
More coarse graining is needed to achieve classical predictability than naive arguments based on the uncertainty principle would suggest. Coarse graining is needed to effect decoherence, and coarse graining beyond that to achieve the inertia necessary to resist the noise that mechanisms of decoherence produce. Sets of histories governed largely by deterministic laws constitute the quasiclassical realm of everyday experience which is an emergent feature of the closed systemÕs initial condition and Hamiltonian. We analyse the the sensitivity of the existence of a quasiclassical realm to the particular form of the initial condition.

\subsection{The Observer Strikes Back \href{http:arxiv.org/abs/1503.07205}{(157)}}
(with Thomas Hertog) In the modern quantum mechanics of cosmology observers are physical systems within the universe. They have no preferred role in the formulation of the theory, nor in its predictions of third person probabilities of what occurs. However, observers return to importance for the prediction of first person probabilities for what we observe of the universe: What is most probable to be observed is not necessarily what is most probable to occur. This essay reviews the basic framework for the computation of first person probabilities in quantum cosmology starting with an analysis of very simple models. It is shown that anthropic selection is automatic in this framework, because there is vanishing probability for us to observe what is where we cannot exist. First person probabilities generally favor larger universes resulting from inflation where there are more places for us to be. In very large universes it is probable that our observational situation is duplicated elsewhere. The calculation of first person probabilities then requires a specification of whether our particular observational situation is assumed to be typical of all the others. It is the combination of the model of the observational situation, including this typicality assumption, and the third person theory which is tested by observation. 

\subsection{Quantum Multiverses \href{http:arxiv.org/abs/1801.08631}{(166)}}
A quantum theory of the universe consists of a theory of its quantum dynamics $(H)$ and a theory of its quantum state $\Psi$. The theory \HPs predicts quantum multiverses in the form of decoherent sets of alternative histories describing the evolution of the universeÕs spacetime geometry and matter content. A small part of one of these histories is observed by us. These consequences follow: (a) The universe generally exhibits different quantum multiverses at different levels and kinds of coarse graining. (b) Quantum multiverses are not a choice or an assumption but are consequences of \HPs or not. (c) Quantum multiverses are generic for simple \HP. (d) Anthropic selection is automatic because observers are physical systems within the universe not somehow outside it. (e) Quantum multiverses can provide different mechanisms for the variation constants in effective theories (like the cosmological constant) enabling anthropic selection. (f) Different levels of coarse grained multiverses provide different routes to calculation as a consequence of decoherence. We support these conclusions by analyzing the quantum multiverses of a variety of quantum cosmological models aimed at the prediction of observable properties of our universe.
In a FAQ we argue that the quantum multiverses of the universe are scientific, real, testable, falsifiable, and similar to those in other areas of science even if they are not directly observable on arbitrarily large scales.

\subsection{Quantum Mechanics at the Planck Scale  \href{http:arxiv.org/abs/gr-qc/9508023}{(123)}}
Usual quantum mechanics requires a fixed, background, spacetime geometry and its associated causal structure. A generalization of the usual theory may therefore be needed at the Planck scale for quantum theories of gravity in which spacetime geometry is a quantum variable. The elements of generalized quantum theory are briefly reviewed and illustrated by generalizations of usual quantum theory that incorporate spacetime alternatives, gauge degrees of freedom, and histories that move forward and backward in time. A generalized quantum framework for cosmological spacetime geometry is sketched. This theory is in fully four-dimensional form and free from the need for a fixed causal structure. Usual quantum mechanics is recovered as an approximation to this more general framework that is appropriate in those situations where spacetime geometry behaves classically.

\subsection{The State of the Universe \href{http:arxiv.org/abs/gr-qc/0209046}{(123)}}
What is the quantum state of the universe? That is the central question of quantum cosmology. This essay describes the place of that quantum state in a final theory governing the regularities exhibited universally by all physical systems in the universe. It is possible that this final theory consists of two parts: (1) a dynamical theory such as superstring theory, and (2) a state of the universe such as HawkingÕs no-boundary wave function. Both are necessary because prediction in quantum mechanics requires both a Hamiltonian and a state. Complete ignorance of the state leads to predictions inconsistent with observation. The simplicity observed in the early universe gives hope that there is a simple, discoverable quantum state of the universe. It may be that, like the dynamical theory, the predictions of the quantum state for late time, low energy observations can be summarized by an effective cosmological theory. That should not obscure the need to provide a fundamental basis for such an effective theory which gives a a unified explanation of its features and is applicable without restrictive assumptions. It could be that there is one principle that determines both the dynamical theory and the quantum state. That would be a truly unified final theory.

\subsection{The Impact  of Cosmology on Quantum Mechanics \href{http:arxiv.org/abs/1901.03933}{(173)}}

When quantum mechanics was developed in the '20s of the last century another revolution in physics was just starting. It began with the discovery that the universe is expanding. For a long time quantum mechanics and cosmology developed  independently of one another. Yet the very discovery of the expansion would eventually draw the two subjects together because it implied the big bang where quantum mechanics was important for cosmology and for understanding and predicting our observations of the universe today. 
Textbook (Copenhagen)  formulations of quantum mechanics are inadequate for cosmology for at least four reasons: 1) They predict  the outcomes of measurements made by observers. But in the very early  universe no measurements were being made and no observers  were around to make them. 2) Observers were outside of the system being measured.  But we are interested in a theory of the whole universe where everything, including observers, are inside. 3) Copenhagen quantum mechanics could not retrodict the past. But retrodicting the past  to understand how the universe began is the main  task of cosmology. 4) Copenhagen quantum mechanics  required a  fixed classical spacetime geometry not least to give meaning to the  time in the Schr\"odinger equation. But in the very early universe spacetime is fluctuating quantum mechanically (quantum gravity) and without definite value. 
A formulation of quantum mechanics general enough for cosmology was started by Everett and developed by many. 
That effort has given us a more general  framework that is adequate for cosmology --- decoherent (or consistent) histories quantum theory in the context of semiclassical quantum gravity.  Copenhagen quantum theory is an approximation to this more general quantum framework that is appropriate for measurement situations.  We discuss whether further generalization may still be required.

\section*{\large Published References}
All of the articles above appear on arXiv at the links shown.  Not all of these have been published in print media.  For those that have, the published references are given below. The numbers in the brackets are the numbers in the author's publication list which is of no interest except to enable reader to 
find the published references from the links above.

\begin{itemize}

\item {68.}  {Computability and Physical Theories} (with R. Geroch), 
{\sl Found. Phys.}, {\bf 16}, 533-550, (1986).

\item {85.}  
Excess Baggage, in {\sl Elementary Particles and the Universe: Essays in
Honor of Murray Gell-Mann}, edited by J.H.~Schwarz, Cambridge University
Press, Cambridge (1991) 1-16, gr-qc/0508001, 

\item{91.} The Quantum Mechanics of Cosmology, in {\sl Quantum
Cosmology and Baby Universes:  Proceedings of the 1989 Jerusalem Winter
School for Theoretical Physics}, eds.~S. Coleman, J.B. Hartle, T. Piran, 
and S. Weinberg, World
Scientific, Singapore (1991) pp. 65-157, arXiv:1805.12246.

\item{98.} The Quantum Mechanics of Closed Systems, in {\sl General
Relativity and Gravitation 1992}, (Proceedings of the 13th Conference on
General Relativity and Gravitation, Cordoba, Argentina, June 28 -- July
4, 1992) ed by   R.J.~Gleiser, C.N.~Kozameh, and O.M.~Moreschi, Institute
of Physics Publishing, Bristol (1993); and in {\sl Directions in
Reltivity, Volume 1, (Essays in Honor of Charles W.~Misner's 60th
Birthday)} ed. by   B.-L.~Hu, M.P.~Ryan, and C.V.~Vishveshwara, Cambridge
University Press, Cambridge (1993); gr-qc/9210006.

\item{107.} Spacetime Quantum Mechanics and the Quantum Mechanics of
Spacetime, in {\sl Gravitation and Quantizations: Proceedings of
the 1992 Les Houches Summer School}, ed.~by B.~Julia \& J.Zinn-Justin,
Les Houches Summer School Proceedings v LVII, North Holland, Amsterdam
(1995); gr-qc/9304006.

\item{108.} Scientific Knowledge from the Perspective of Quantum
Cosmology, in the {\sl Boundaries and Barriers: On the Limits of
Scientific Knowledge} (1995 Abisko Conference), 
ed. by  J.L.~Casti and A.~Karlqvist,
Addison-Wesley Publishing Co., Reading, MA
(1996); gr-qc/9601046.

\item{112.} Sources of Predictability, in {\sl Complexity}, {\bf 3}, 22-25,
1997; gr-qc/9701027. 

\item{113.} Quantum Cosmology: Problems for the 21st Century, in {\sl
Physics in the 21st Century: Proceedings of the 11th Nishinomiya-Yukawa
Symposium}, Nishinomiya, Hyogo, Japan, ed. by  K.~Kikkawa, H.~Kunitomo, and
H.~Ohtsubo, World Scientific, Singapore (1997); gr-qc/9701022.

\item{117.}
Quantum Pasts and the Utility of History,
in the {\sl The Proceedings of the
Nobel Symposium: Modern Studies of Basic Quantum Concepts and
Phenomena},
Gimo, Sweden, June 13-17, 1997, {\sl Physica
Scripta}, {\bf T76}, 67--77 (1998); gr-qc/9712001.

\item{125.} Theories of Everything and Hawking's Wave Function of the
Universe, in {\sl The Future of
Theoretical Physics and Cosmology: Celebrating Stephen Hawking 
60$^{th}$ Birthday},
ed.~by G.W.~Gibbons, E.P.S.~Shellard, and S.J.~Ranken,
(Cambridge University Press, Cambridge UK, 2003); gr-qc/0209047.

\item{131.} The Physics of Now, {\sl Am. J. Phys.}, {\bf 73}, 101-109
(2005), gr-qc/0403001.


\item{136.}  Quantum Physics and Human Language, {\sl  J. Phys. A: Math. Theor.},  {\bf 40}, 3101-3121 (2007), quant-ph/0610131.  

\item{137.}  Are We Typical? (with M. Srednicki), {\sl Phys. Rev. D}, {\bf 75}, 123523 (2007), arXiv:0704:2530. 

\item{144}  Science in a Very Large Universe (w. M. Srednicki), Phys. Rev. D, 81 123524 (2010), arXiv:0906.0042, a summary of this work with different emphases is in: The Xerographic Distribution: Scientific Reasoning in a Large Universe, arXiv:1004.3816.


\end{itemize}

\end{document}